\documentclass[11pt]{article}
\usepackage{preprint}
\usepackage{epsfig}

\usepackage{graphicx}
\usepackage{xspace}
\let \IG \includegraphics

\newcommand{\sql}{standard quantum limit}
\newcommand{\eom}{electro-optic modulator}

\newcommand{\mFSR}{\ensuremath\rm FSR}
\newcommand{\eg}{\mbox{e.\,g.}\xspace}
\newcommand{\shot}{shot noise}

\newcommand{\shotH}{shot-noise}

\newcommand{\Geo}{GEO\,600}

\newcommand{\sr}{signal recycling}

\newcommand{\Por}{Power recycling}
\newcommand{\por}{power recycling}

\newcommand{\pHr}{power-recycling}

\newcommand{\prc}{power-re\-cycling cavity}

\newcommand{\src}{signal-re\-cycling cavity}

\newcommand{\pom}{power-recycling mirror}

\newcommand{\sIm}{signal-recycling mirror}

\newcommand{\pod}{power-recycled}

\newcommand{\dud}{dual-recycled}
\newcommand{\Dur}{Dual recycling}
\newcommand{\dur}{dual recycling}

\newcommand{\Mi}{Michelson interferometer}

\newcommand{\mc}{mode cleaner}

\newcommand{\gHw}{gravi\-ta\-tion\-al-wave}

\newcommand{\PDH}{Pound-Dre\-ver-Hall}
\newcommand{\FSR}{free spectral range}
\newcommand{\pd}{photo diode}

\newcommand{\bs}{beam splitter}

\newcommand{\mFig}[1]{Figure~\ref{#1}}

\newcommand{\SSm}{\scriptscriptstyle\rm}
\newcommand{\amaldi}{Class.~Quantum Grav.~{\bf 19} (2002)}

\newcommand{\adf}{A.~Freise}

\def\be{\begin{equation}}
\def\ee{\end{equation}}
\def\bea{\begin{eqnarray}}
\def\eea{\end{eqnarray}}

\begin{document}
\renewcommand{\topfraction}{1.0}
\renewcommand{\bottomfraction}{1.0}
\renewcommand{\textfraction}{0.0}
\renewcommand{\floatpagefraction}{.1}

\vspace*{4cm}
\title{\Dur\ for \Geo}

\author{A.~Freise for the \Geo\ team}

\address{Albert-Einstein-Institut Hannover,\\
Callinstr. 38, 30167 Hannover, Germany}

\maketitle\abstracts{
\Dur\ is the combination of \emph{\sr} and \emph{\por}; both optical techniques 
improve the \shotH-limited sensitivity of interferometric \gHw\ detectors. 
In addition, \sr\ can reduce the loss of light power due to imperfect interference 
and allows, in principle, to beat the
\sql. The interferometric
\gHw\ detector \Geo\ is the first detector to use \sr. We have recently equipped
the detector with a \sIm\ with a transmittance of $1\%$. 
In this paper, we present details of the detector commissioning
and the first locks of the \dud\ interferometer.
}

\section{Introduction}
\begin{figure}[t]
\begin{center}
\IG [scale=.28] {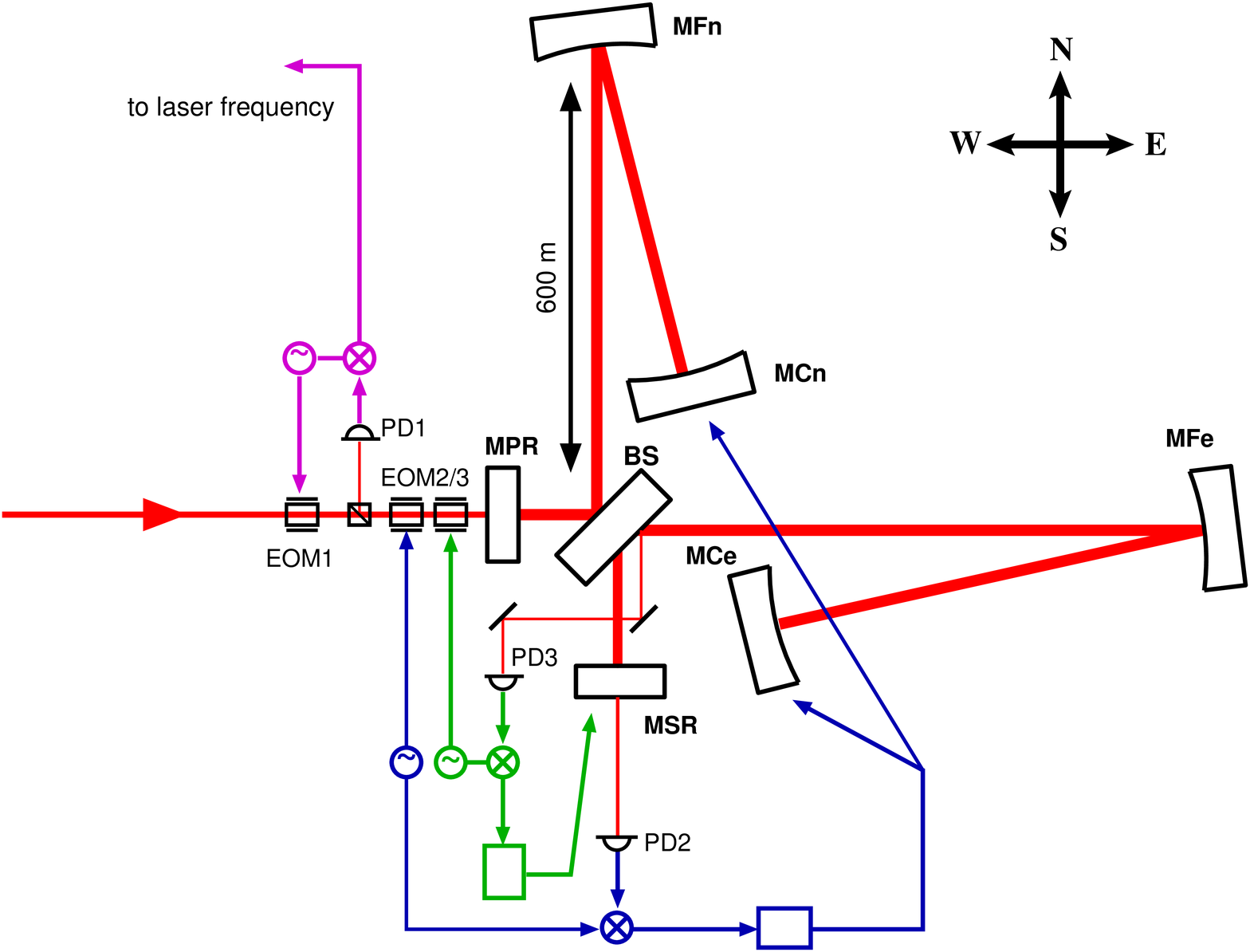}
\end{center}
\caption{\label{fig:mi_sch1}Simplified schematic of the \Geo\ \Mi . The beam enters the interferometer 
from the west through the
\emph{\pom } (MPR) and is split at the main \bs\ (BS).  Each arm of the interferometer is folded 
once by the \emph{folding mirrors} (MFn and MFe) so that each arm has a round-trip length
of 2400\,m. The folding is vertical but illustrated above as a
horizontal setup. The \emph{end mirrors} are MCn and MCe. 
Beams leave the \Mi\ at the west and the south ports. The output used for signal detection 
is the south port with the \sIm\ (MSR). The three longitudinal degrees of freedom are controlled
by three feedback loops, for which the laser light is phase-modulated
before entering the interferometer. One pair of control sidebands 
({\rm EOM1}, frequency $f_{\rm PR}\approx37.2$\,MHz)
is used for controlling the \prc\ via a standard \PDH\ scheme (PR loop), which is 
part of the laser frequency stabilisation system.
Two further pairs of control sidebands (at frequencies $f_{\rm MI}\approx14.9$\,MHz and 
$f_{\rm SR}\approx9$\,MHz) are created with {\rm EOM2} and {\rm EOM3}
to be resonant in the recycled interferometer. The error 
signal for controlling the operating point of the \Mi\ (MI loop) is obtained by demodulating the
signal of the main \pd\ {\rm PD2} in the south port (at $f_{\rm MI}$). Another \pd\ {\rm PD3} 
is used to detect the
light reflected by the AR coating of the \bs; the \pd\ signal is demodulated 
at $f_{\rm SR}$ to generate the error signal for controlling the length of the
\src\ (SR loop). \hfill \ }
\end{figure}

To date, all interferometric \gHw\ detectors are basically 
Michelson interferometers optimised to measure
tiny differential phase modulations of the light in the two arms.
One of the main noise sources of these instruments is the photon
\shot. To reduce the effect of \shot, the established 
technique of power recycling is used by all interferometric detectors.
\Por\ allows to maximise the light power circulating in the interferometer
arms without reducing the bandwidth of the detector or installing
a larger laser.
The sensitivity can be improved further by optimising the 
signal storage time, i.e.\ the average time for which the 
phase modulation sidebands induced by the gravitational wave
are stored in the interferometer.
For this purpose several advanced techniques have been proposed.
For example, \sr\ in combination  with \por\ was proposed by Meers~\cite{meers:dr,strain91} 
and is called {\em \dur}. 
It will be implemented in the \Geo\ detector from the beginning on and
is planned for the second generation of interferometric
detectors. \Geo\ is a British-German interferometric detector with 600\,m long arms.
\mFig{fig:mi_sch1} shows a schematic diagram of the \Geo\ \Mi.
The construction of the detector is complete and the \Mi\ is currently 
being commissioned \cite{hal}.
\vspace{\baselineskip}

The \pHr\ technique exploits the fact that at the operating point
\emph{dark fringe} the \Mi\ (MI) 
behaves like
a mirror for the injected light; the \pom\ (MPR) recycles
the light from the bright fringe in the west port, forming
a cavity with the \Mi , the \emph{\prc } (PRC). This cavity is kept on resonance by a \PDH\
method to resonantly enhance the carrier light inside the cavity. 
The maximum possible power enhancement is limited by losses in the interferometer.
Assuming a \prc\ with $T_{\SSm MPR},T_{\SSm L}\ll1$ we can write:
\begin{equation}
{P_{\rm cav}}/{P_{\rm in}}=G_{\rm PR}\approx{4 T_{\SSm MPR}}/{\left(T_{\SSm MPR}+T_{\SSm L}\right)^2}
\end{equation}
where $T_{\SSm L}$ is the power loss factor of one complete round-trip
in the PRC and $T_{\SSm MPR}$ the power transmittance of MPR. The ratio $G_{\rm PR}$ 
between the injected power and the intra-cavity power is called \emph{\pHr\ gain}. 
In the \Geo\ detector, the
main origin of power loss is the imperfect interference at the
\bs\ due to a mismatch of the beam phase fronts. Currently, this is caused mainly by a 
mismatch in the radii of curvature of the end mirrors. In the final state the
effect will be dominated by the thermal lens in the \bs\ substrate. 
\enlargethispage*{\baselineskip}

A passing gravitational 
wave (or an equivalent differential motion of the
end mirrors $\rm MCn$ and $\rm MCe$) will modulate the phase of the
light in the interferometer arms and thus create `signal sidebands' that
are not directed to the west but to the south port. 
The power of the signal sidebands on the \pd\ can be enhanced independently of the carrier
light. The \sIm\ (MSR) in the south 
port reflects the signal sidebands back into the \Mi . Again, the recycling
mirror forms a cavity together with the \Mi, the \src\ (SRC), which is
tuned to resonantly enhance the signal sidebands. 
The maximum power enhancement of the signal sidebands on the photo detector, outside the SRC  
(over-coupled and $T_{\SSm MSR}\ll1$), is given
by:
\begin{equation}
G_{\SSm SR}={4}/{T_{\SSm MSR }}
\end{equation}

As the \sIm\ closes the south port of the interferometer, it can considerably reduce
the amount of carrier light power being lost through that port. 
FFT simulations have
shown that the residual light power at the carrier frequency 
transmitted into the south port can be found almost
entirely in second-order TEM modes (with respect to the eigenmode of the
PRC). With MSR in place the SRC is (by design) not resonant for second-order TEM
modes. Thus, these light fields are suppressed and the power loss from the PRC is
reduced. This effect is also called \emph{mode healing}. The FFT simulations have been used
to predict the exact amount of the expected mode-healing effect\cite{ros}. 
In the case of the current \Geo\ detector and assuming negligible other losses 
inside the interferometer, we expect that the \pHr\ gain 
can be approximated as:
\begin{equation}
G_{\rm PR}\approx{4 T_{\SSm MPR}}/{\left(T_{\SSm MPR}+T_{\SSm MSR}T_{\SSm l}\right)^2}
\end{equation}
with $T_{\SSm l}$ as the power loss due to the phase front mismatch at the \bs.
The mode healing effect also works for the SRC, so that MPR
ensures a high finesse of the SRC.

\section{Control of the \dud\ interferometer}
\begin{figure}[th]
\begin{center}
\IG [bb=0 0 335 155,scale=1.28] {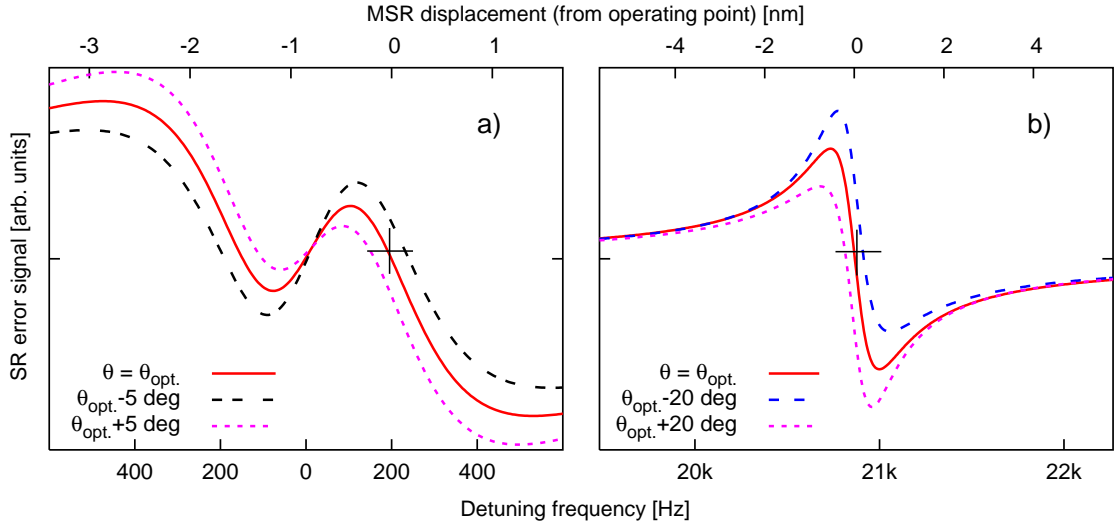}
\end{center}
\caption{\label{fig:sr_west}Simulated SR error signal for a) 
a typical detuning of 200\,Hz that would be optimal for broadband
operation and b) a large detuning of $\approx20$\, kHz (the operating 
points are indicated with a cross).
Both graphs show the error signal for an optimised demodulation frequency and phase and 
for slightly different demodulation phases. 
It can be seen that
the signal for 200\,Hz detuning is suitable for controlling the MSR but not optimal for
the lock acquisition process.
The nearby zero crossing at 0\,Hz requires that the control loop reduces any residual 
motion of MSR to well below one nm throughout all stages of acquisition.
In addition, the signal is very sensitive to the demodulation phase.
Using large detuning provides a symmetric signal and
the residual motion of the MSR may be up to a few nm without 
causing the lock acquisition to fail. In addition, the dependency on the demodulation phase
is also reduced. Both effects relax the requirements for the control loop and result in a
more robust lock acquisition scheme. \hfill\ }
\end{figure}
The \dud\ \Mi\ has three longitudinal degrees of freedom: (1) the length of the \prc, (2) the 
operating point of the MI (differential arm length),
and (3) the length of
the \src\ (see \mFig{fig:mi_sch1}).
The length of the \prc\ is used as reference for the laser
frequency stabilisation \cite{adf}. 
Error signals for the two remaining degrees of freedom 
are obtained with the so-called \emph{Schnupp modulation} technique
(also known as `frontal modulation' or `pre-modulation'):
The laser light is
phase-modulated (at RF) by an \eom\ before it enters the 
interferometer. This creates two control sidebands that are
injected into the interferometer together with the carrier. 
The beat note between the control sidebands and
the carrier light leaking out of the \Mi\ generates a signal proportional
to the \Mi's deviation from the dark fringe.
The Schnupp modulation method can also be used to generate error signals with
respect to other degrees of freedom. In particular, we use 
an additional Schnupp modulation at a different modulation
frequency to generate a control signal for the SRC length.

With the \prc\ and the \Mi\ being at their nominal operating points,
the resonance condition inside the \src\ is determined by the
position of MSR along the optical axis. 
Each microscopic position of MSR corresponds to
a different Fourier frequency of maximum enhancement.
Therefore, a change in the position of MSR is also
called \emph{tuning} or \emph{detuning} of the mirror and thus
of the detector:
\begin{equation}
\delta x_{\SSm MSR}=\frac{\lambda}{2}~\frac{f_{\rm sig}}{\mFSR_{\SSm SRC}}=4.23\,{\rm pm}\cdot\left( \frac{f_{\rm sig}}{1\,{\rm Hz}}\right)
\end{equation}
with $\delta x_{\SSm MSR}$ as the detuning of MSR, 
$\lambda$ the carrier wavelength, $f_{\rm sig}$ the signal or \emph{detuning frequency} and
$\mFSR_{\SSm SRC}$ the \FSR\ of the SRC.
The \Geo\ detector can easily be tuned to a user-defined Fourier frequency
by adjusting the modulation frequency and
the demodulation phase of the SR loop and the gain of the
electronic servos. We expect to continuously change the tuning 
of the detector during normal operation
without losing lock of the system.
A description of the general control concept is given elsewhere \cite{ghhdr,adf_dr,ghh02}.
\enlargethispage*{\baselineskip}

\section{Commissioning of the detector}
During the commissioning of the \Geo\ detector we have first set up
a \pod\ MI using a \pom\
with $T_{\rm MPR}\approx1.4\%$ transmittance. The laser
light power was reduced so that 1\,W was injected into the main interferometer.
The control systems of the \pod\ MI have been proven to work very reliably;  
during the S1 science run in summer 2002\cite{mh}, for example, the longest continuous lock of
all subsystems lasted over 120 hours and the overall duty cycle was $>98\%$.
Recently, we have equipped the interferometer with a \sIm\ having a power transmittance 
of $T_{\rm MSR}=1\%$, which yields a detector bandwidth of $\approx 200\,$Hz. 
First locks
of the \dud\ system, up to 15 minutes long,
have been achieved. In addition, the interferometer
can also be used in a \pod\ mode, simply by misaligning MSR.
\begin{figure}[th]
\begin{center}
\begin{minipage}[c]{0.52\linewidth}
\IG [bb=0 0 265 250, width=\linewidth] {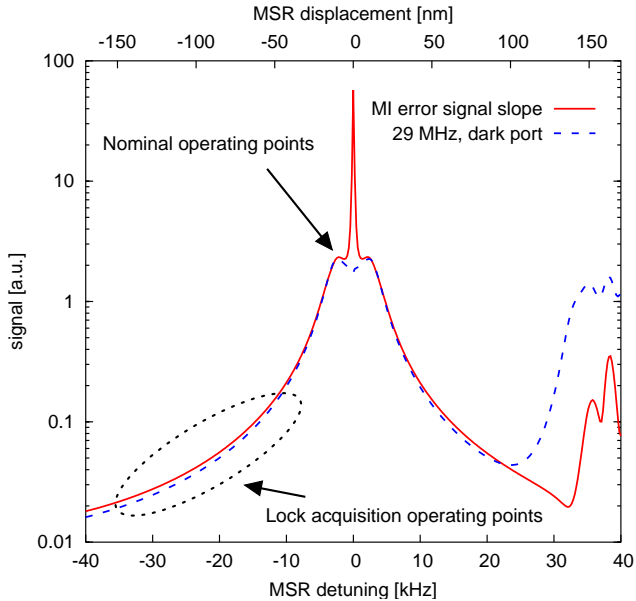}
\end{minipage}\hfill \raisebox{.1cm}{\begin{minipage}[c]{0.44\linewidth}
\caption{\label{fig:normal}Simulated error signal slope of the MI control signal
and a possible normalisation signal.
The nominal operating points are at detunings between $0$\,Hz and
a few kHz. During lock acquisition we expect to use largely detuned
operating points in a region between $-10$\,kHz and $-$35\,kHz (the complex 
structure for detunings $>20$\,kHz, a
resonance of the second-order TEM modes in the SRC, will probably prevent
lock acquisition for large positive detunings). The gain of the MI loop changes
by more than three orders of magnitude when the MSR tuning is changed.
A comparison of the available output signals showed that the
power of the sidebands at $\approx29$\,MHz (the first harmonic of
the MI control sidebands) detected in the south port 
can be used to determine the optical gain, except for very small detunings. 
\hfill \ }
\end{minipage}}
\end{center} 
\end{figure}
\enlargethispage*{\baselineskip}

Switching from \por\ to \dur\ requires a number of alterations to the
control system. The modulation frequencies for the MI 
loop and the SR loop now have to be resonant in the 4-mirror resonator formed by 
MPR, MI and MSR.
These frequencies are
chosen to approximately match a multiple of the free
spectral range of the \prc:
\begin{equation}
f_{\rm MI}\approx14.9\,{\rm MHz}\,\approx119\cdot \mFSR_{\SSm PRC},\qquad\qquad
f_{\rm SR}\approx9\,{\rm MHz}\,\approx72\cdot \mFSR_{\SSm PRC}
\end{equation}
An alternating sequence 
of measurements and simulations was used to determine the state of
the detector and set the exact modulation frequencies and demodulation phases. 

The expected SR error signals at nominal operating points have turned out not to be 
well suited for lock acquisition, see \mFig{fig:sr_west}. 
More symmetric and robust error signals can be
generated when the detector is largely detuned to
frequencies between, \eg, 10\,kHz and 35\,kHz. 
Therefore we decided to use such largely detuned operating points for the
initial lock acquisition and then tune the detector while remaining in lock to its nominal
operating point. For this purpose we plan to use an \emph{automatic gain control} which
takes the power of the 29\,MHz sidebands as a measure of the optical gain
of the MI loop, see \mFig{fig:normal}.

The three control loops (PRC, MI, SR) are subject to a natural
hierarchy: First, a proper error signal for the MI exists only after 
the PRC has been locked to the laser light, whereas the PRC can be locked for almost
all states of the MI by virtue of the asymmetric reflectance of the
\bs. Second, a SR error signal can be obtained only 
after the MI is locked to the dark fringe whereas the position
of MSR affects mainly the gain the MI error signal. These gain variations can be
compensated using an appropriate normalisation signal (see below).
Therefore, the lock acquisition scheme for \Geo\ has been designed as follows:

When the \mc s are locked and the laser light is present
at the \pom, the PR control closes the servo loop when the 
cavity is close to resonance. The optical gain of the PR loop depends on
the state of the MI. In order to compensate for the gain changes, an automatic
gain control is used: The PR error signal is divided by a signal proportional
to the light power inside the cavity.
When a PR lock is achieved, the MI control waits for the MI to
pass through a dark fringe. 
A micro-controller is used to analyse several
interferometer channels and to close the MI servo at a dark fringe if the differential
mirror motion at that dark fringe is slow enough to be stopped by
the force of the actuators.
As soon as the MI is locked, the gain of this loop is increased to harden
the control system against the influence of the still moving MSR. Finally, the SR 
control is switched on and closes the servo close to a resonance of the
SRC.

The speed of the SR lock acquisition can be increased by using the following trick:
The range of possible large detunings is relatively wide. Instead of setting
the servo to one particular operating
point (by adjusting the modulation frequency and phase),
we use a micro controller to quickly sweep the operating point (by
sweeping modulation frequency and phase). If the MSR position is somewhere
in the required range, the controller will find it, determined by a well-defined 
zero crossing of the error signal, and thus set the nominal operating exactly
to that position and close the servo loop.
\enlargethispage*{\baselineskip}
\begin{figure}[th]
\begin{center}
\IG [bb=10 0 350 205, scale=1.22] {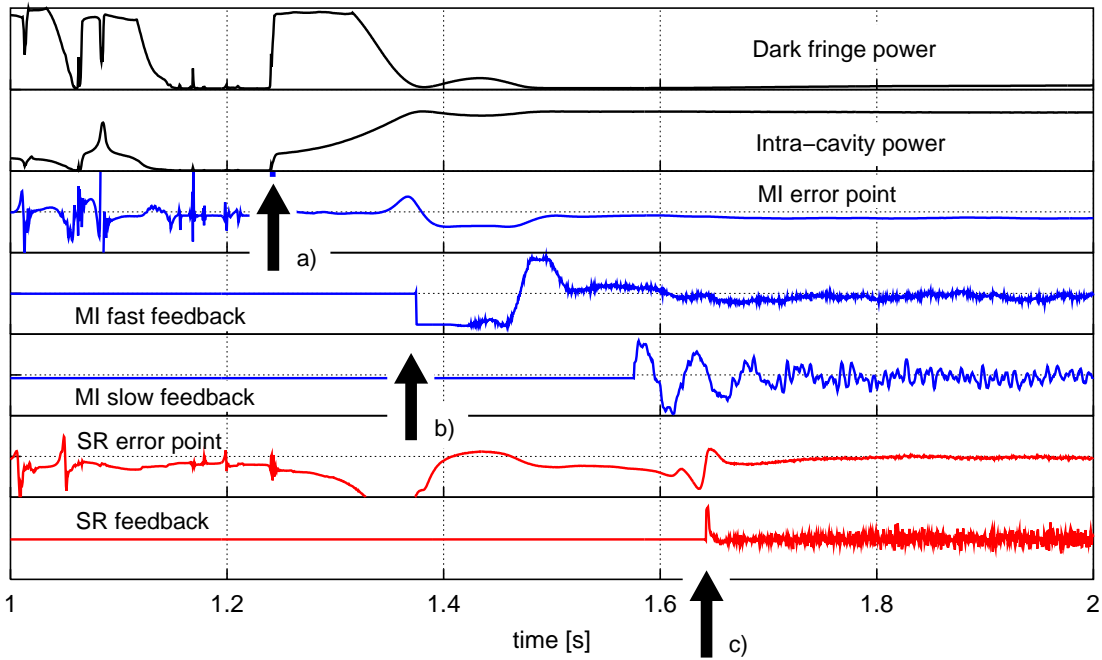}\\ 
\end{center}
\caption{\label{fig:lock}Measured time trace of interferometer channels during a 
full lock acquisition of the \dud\ detector. The arrows indicate the time
when the respective locks are achieved: a) the PRC is locked, light is injected into the
MI which is, at that moment, not at the dark fringe; thus, the intra-cavity power 
is comparably low and the south port is bright. A tenth of a second later 
the MI passes through a dark fringe:
the light power in the south port is reduced whilst the intra-cavity power builds up.
Near the dark fringe the MI error point signal becomes visible. At its zero crossing
the MI control loop is engaged, indicated by arrow b). First, the fast feedback
(using an electro-static actuation on the test mass) is switched on. Some time later 
a slow feedback (via magnet-coil actuators at the intermediate mass) is switched on to 
increase the loop gain. In the following, the MI remains 
at the dark fringe while residual motion of the end mirrors is damped out. 
Only a few tenths of a second later the SRC passes through 
the SR operating point. The SR control loop is closed at 
the zero crossing of the SR error signal, marked by arrow c), 
 and the SRC is locked. \hfill\ }
\end{figure}

Using the method described above we have achieved first locks of the
\dud\ detector.
\mFig{fig:lock} shows a measured time trace of several interferometer signals
during a full lock acquisition. The acquisition time in this example was about
1 second. Typical durations of locks of the \dud\ system currently range up to
15 minutes. We have also reached $\approx300$\,W 
of power at the \bs\ (compared to 250\,W without SR), 
which corresponds to the expected
power increase due to the mode-healing effect.
\vspace{-3mm}

\section*{Acknowledgements}
The author would like to thank PPARC in the UK, the BMBF and the state of Lower Saxony in Germany.

\end{document}